\documentclass{egpubl}
\usepackage{eurovis2025}
\CGFccby
\usepackage[T1]{fontenc}
\usepackage{dfadobe}  

\usepackage{cite}
\BibtexOrBiblatex

\electronicVersion
\PrintedOrElectronic
\ifpdf \usepackage[pdftex]{graphicx} \pdfcompresslevel=9
\else \usepackage[dvips]{graphicx} \fi

\usepackage{egweblnk}
\usepackage{xspace}
\usepackage{relsize}
\def\code#1{{{\relsize{-1}\texttt{#1}}}\xspace}
\usepackage{amssymb}
\usepackage{relsize}
\usepackage{booktabs}
\usepackage{tabu}
\usepackage{multirow}
\usepackage[capitalise]{cleveref}
\usepackage{tikz}
\usepackage{listings}
\usepackage{stackengine}
\usepackage{algorithm}
\usepackage{algpseudocode}
\usepackage{makecell}
\usepackage{url}
\usepackage{breakurl}

\definecolor{ForestGreen}{HTML}{009B55}

\usepackage{listings}
\newif\ifsubmission
\title[Data Parallel Visualization and Rendering on the RAMSES Supercomputer with ANARI]{Data Parallel Visualization and Rendering on the\\RAMSES Supercomputer with ANARI}
\author[S.~Zellmann]
{\parbox{\textwidth}{\vspace{-4em}\centering
    Stefan~Zellmann\orcid{0000-0003-2880-9090}}}
\definecolor{darkgreen}{rgb}{0,0.5,0}
\definecolor{midgreen}{rgb}{0,0.6,0}
\definecolor{lightgreen}{rgb}{0,0.8,0}
\definecolor{darkred}{rgb}{0.6,0,0}

\newif\ifdiff
\ifdiff
\usepackage{soul}

\def\removed#1{\textrm{\color{red}\st{#1}}}
\else

\def\removed#1{}
\fi

\teaser{
  \vspace{-4em}
 \includegraphics[width=0.16\linewidth]{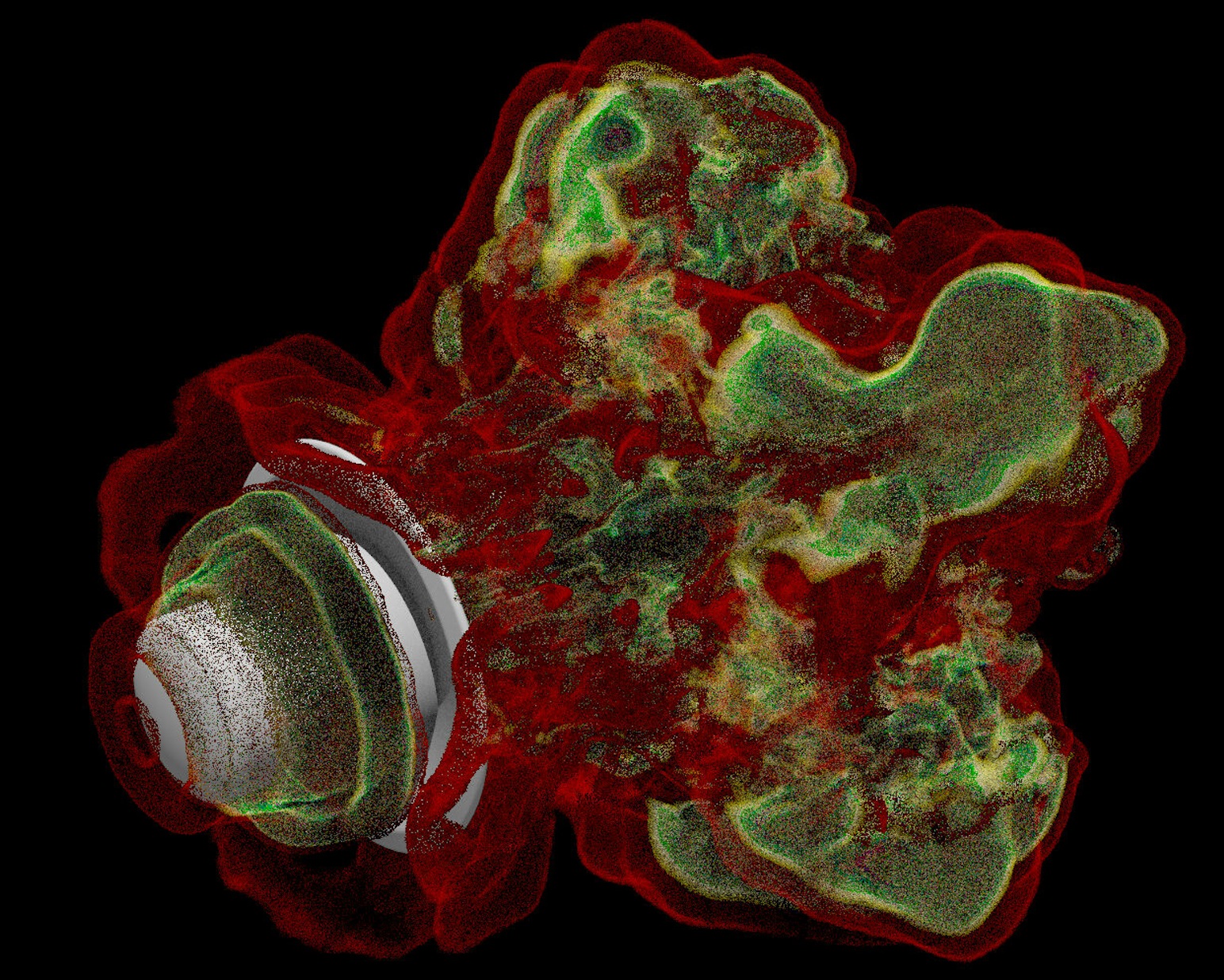}
 \includegraphics[width=0.16\linewidth]{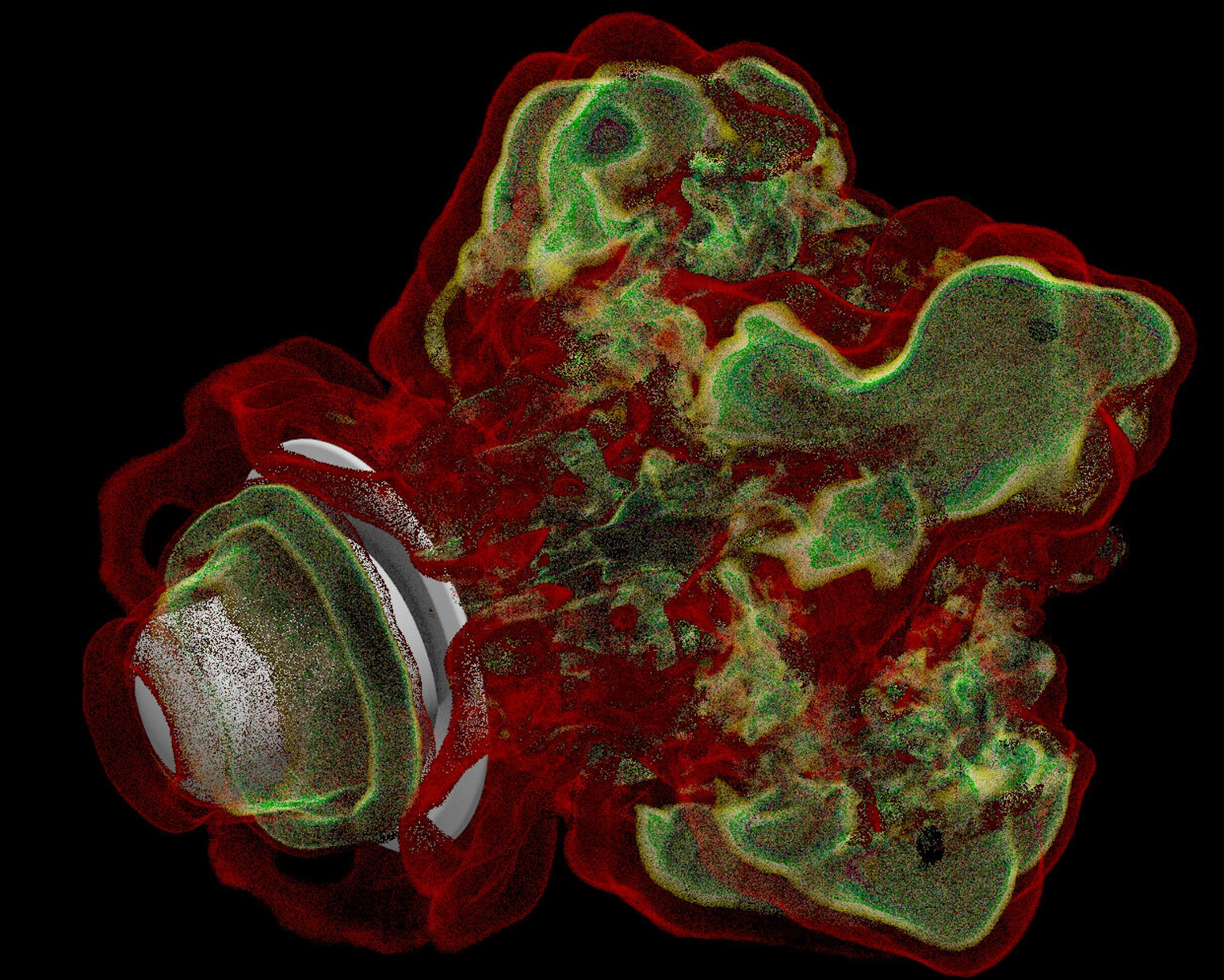}
 \includegraphics[width=0.16\linewidth]{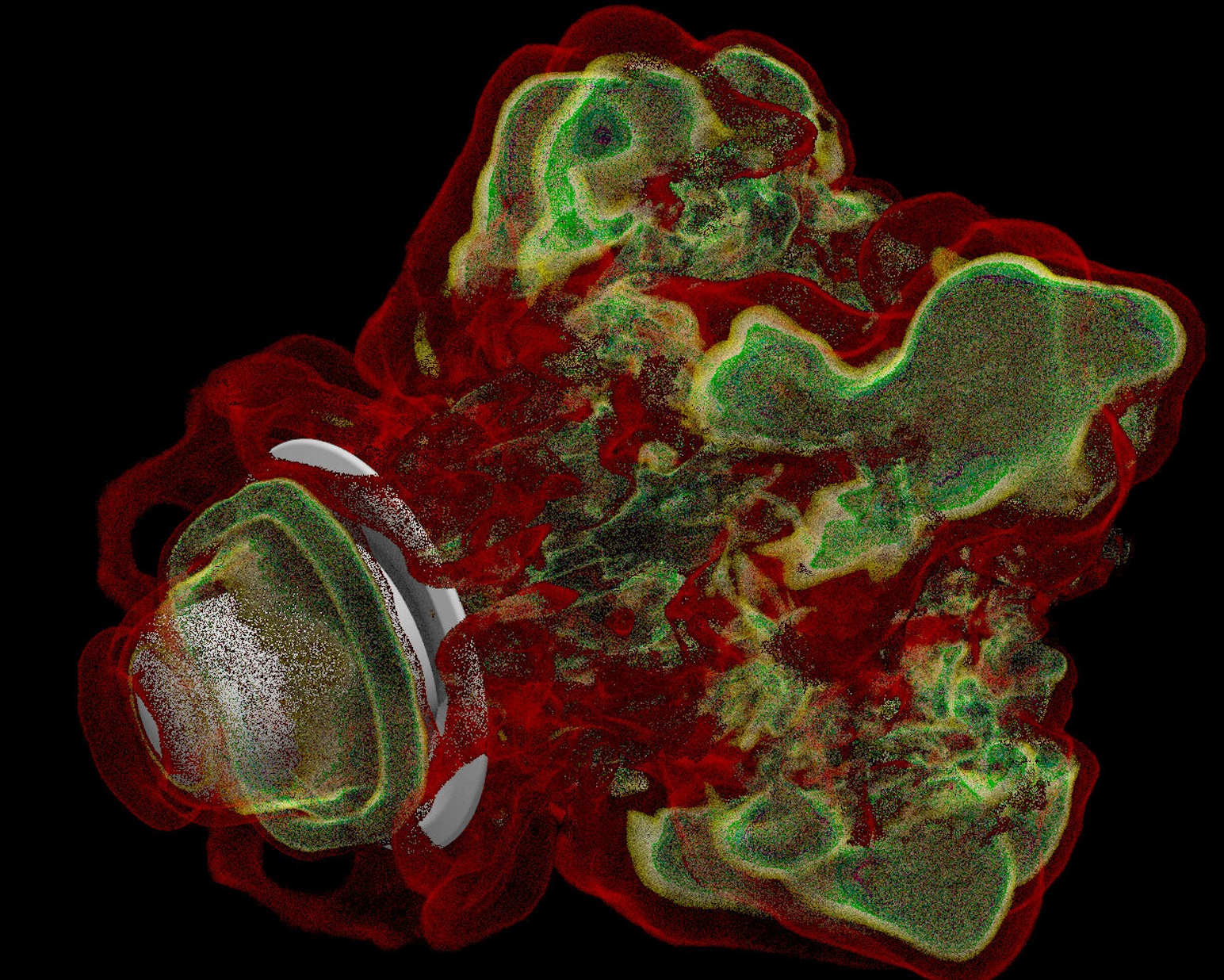}
 \includegraphics[width=0.16\linewidth]{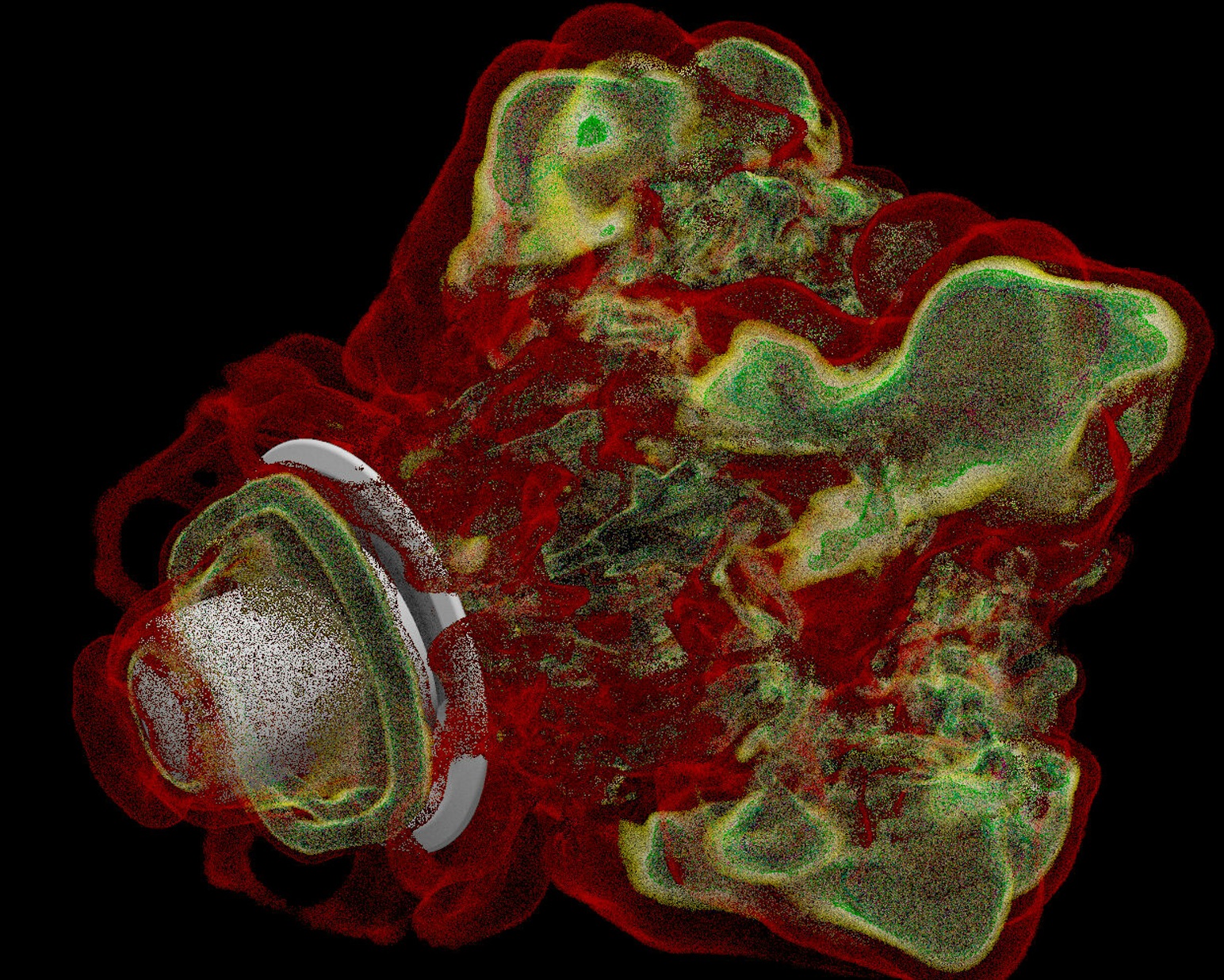}
 \includegraphics[width=0.16\linewidth]{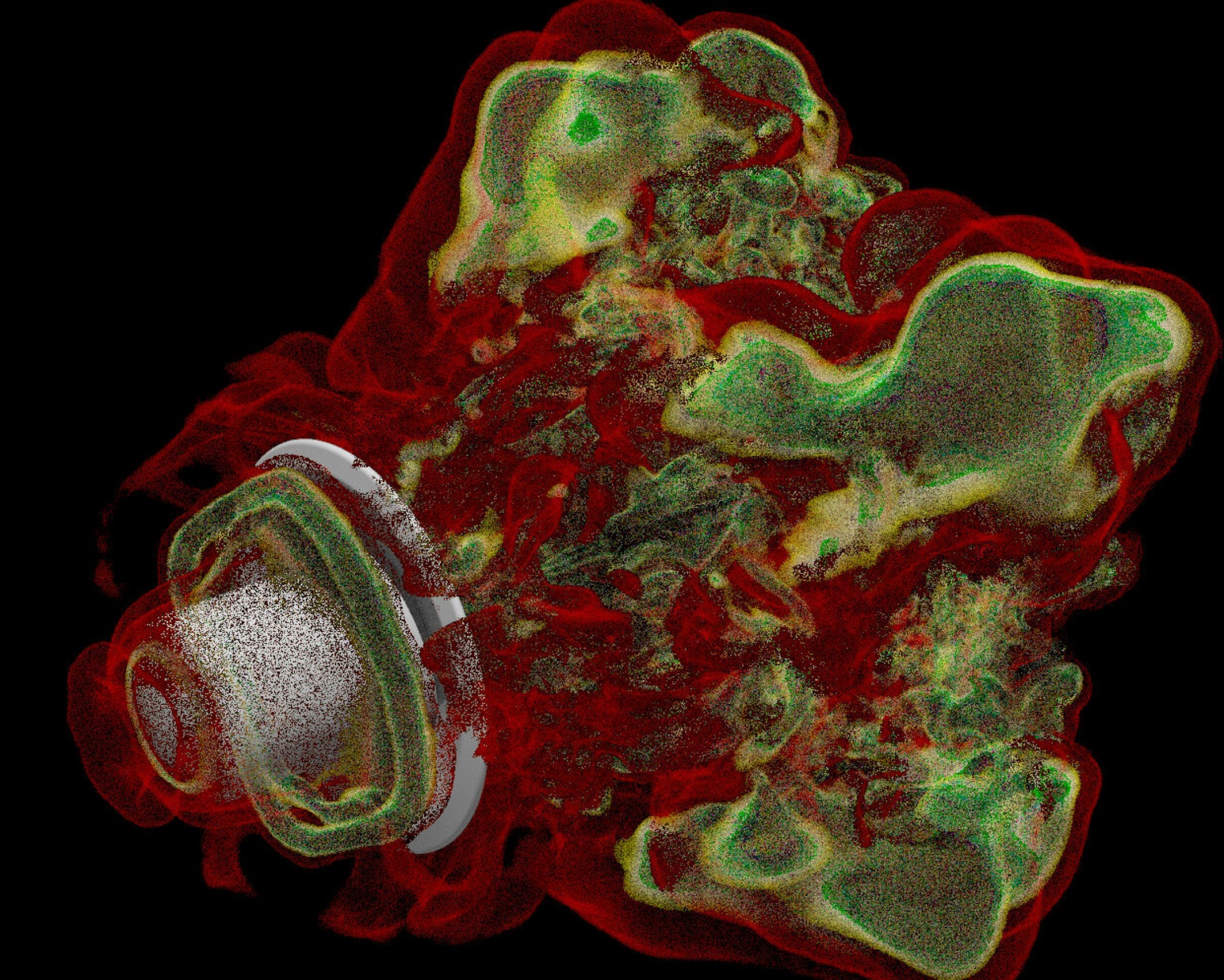}
 \includegraphics[width=0.16\linewidth]{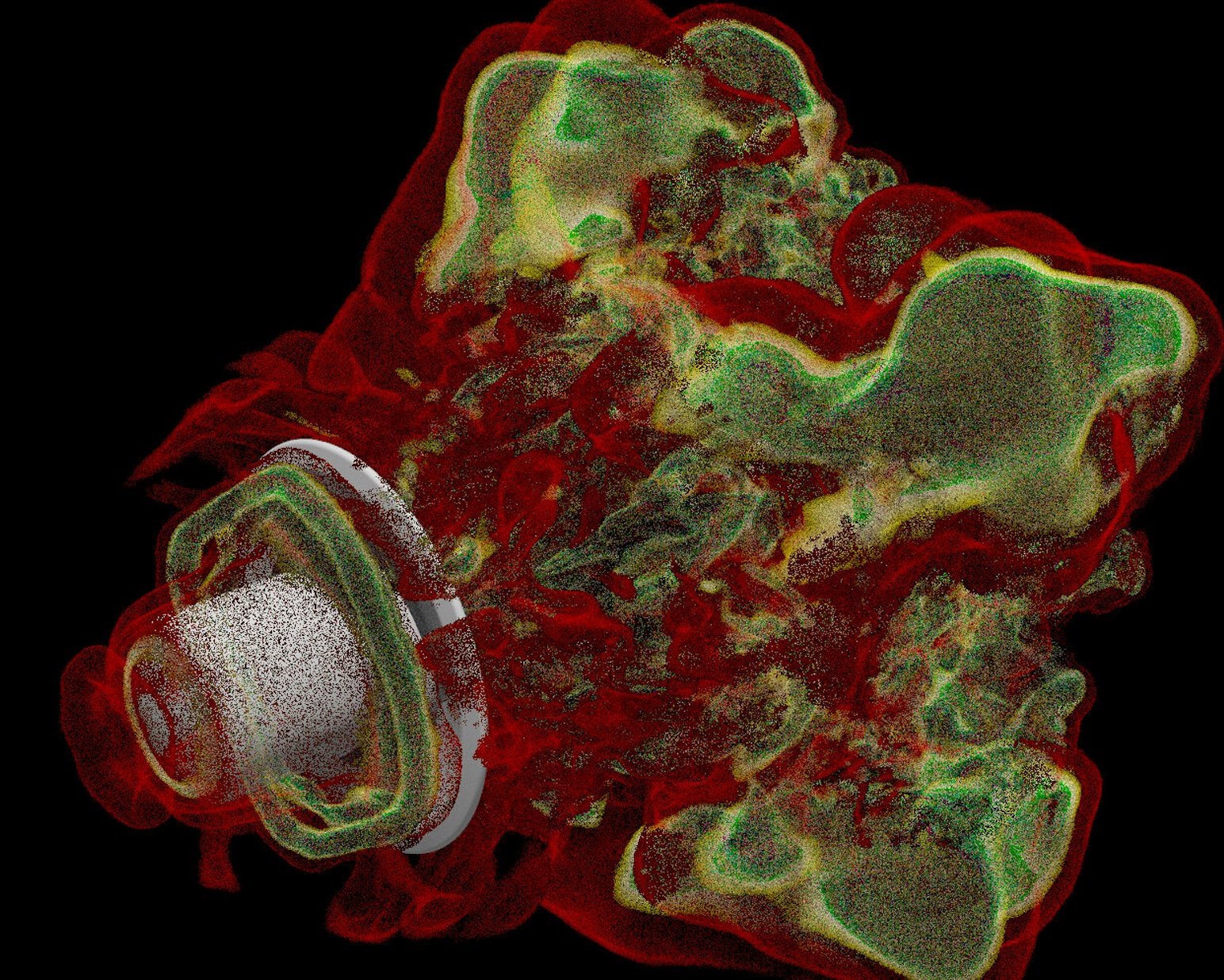}
 \centering
 \vspace{-0em}
 \caption{\label{fig:teaser}%
 Six time steps of a large-scale scientific visualization of a computational
 fluid dynamic simulation---conducted by NASA for the Mars Lander atmospheric
 entry~\cite{Jones2019marslander}. This simulation was conducted on Summit on
 over 500 GPUs. The visualization was rerun on RAMSES using 36 NVIDIA~H100
 GPUs. The data set is partitioned into over 500 parts and comprises multiple
 time steps and field variables. Distributed rendering with ANARI allows to
 interactively  visualize such data in-situ on the HPC system using ray
 tracing. We demonstrate a practical implementation of this on the RAMSES
 supercomputer at the University of Cologne.
 }
\label{fig:teaser}
}

\begin{document}

\maketitle
%-------------------------------------------------------------------------
\begin{abstract}
3D visualization and rendering in HPC are very heterogenous applications,
though fundamentally the tasks involved are well-defined and do not differ much
from application to application. The Khronos Group's ANARI standard seeks to
consolidate 3D rendering across sci-vis applications. This paper makes an
effort to convey challenges of 3D rendering and visualization with ANARI in the
context of HPC, where the data does not fit within a single node or GPU but
must be distributed. It also provides a gentle introduction to parallel
rendering concepts and challenges to practitioners from the field of HPC in
general. Finally, we present a case study showcasing data parallel rendering
on the new supercomputer RAMSES at the University of Cologne.
\end{abstract}

\section{Introduction}
\label{sec:intro}
Scientific visualization (sci-vis) applications implement a variety of
post-processing algorithms on scientific, three-dimensional data. Some of those
algorithms are concerned with filtering the data to extract features of
interest, to map such features to colors, different shapes, etc.; one
particularly important algorithms is rendering that takes the filtered and
mapped data and turns it into a 2D raster image that is interactively refreshed
on user interaction, facilitating exploration.

A multitude of sci-vis software systems targeted at high performance computing
(HPC) applications exists. Some of these software systems are quite general,
such as ParaView~\cite{paraview} or VisIt~\cite{visit} (both of which
internally use the VTK library~\cite{vtkBook} to implement vis algorithms).
Other sci-vis apps tackle very specific HPC problems~\cite{ovito,vmd}.
Rendering is often implemented using rasterization and OpenGL.

Large-scale scientific rendering of simulation data on HPC and supercomputing
systems is often implemented using ray tracing these days. HPC systems provide
limited resources for hardware rasterization rendering with OpenGL and similar
APIs. The hardware vendors provide optimized ray tracing kernel frameworks to
accelerate ray/object intersection tests; noteworthy examples are Intel's
Embree~\cite{embree} and NVIDIA's OptiX~\cite{optix}.

This gives rise to a multitude of low-level rendering APIs for rasterization
and ray tracing, and vis apps that implement their own \emph{renderers} on top.
3D rendering in general is rather well-defined: the input is a structured
description of the 3D geometry or volumetric data plus color or normal maps,
light sources, and a virtual camera description to generate the image
from. All these entities can change over time, though depending on the
application, certain entities, such as the camera, are more likely to change
than others (e.g., the geometry or volume elements). In principle, from the
application's side, the task of generating 2D imagery from such a structured
description can be treated as a black box, given a couple of parameters and
settings for image quality or interactivity.

Yet, the default approach implementing renderers so far has been for the app to
implement a custom rendering subsystem that is deeply integrated into the vis
app itself; VTK for example has a custom OpenGL renderer; similar renderers are
also found in virtually any other vis app. OSPRay~\cite{ospray} and
VisRTX~\cite{visrtx} are vendor specific rendering libraries on top of Embree
and OptiX.

The Khronos Group's ANARI standard~\cite{anari} is an effort to consolidate
those rendering submodules into rendering back-ends with a well-defined API.
ANARI draws inspiration from both OSPRay and VisRTX, and seeks to hide the
specifics of 3D rendering behind that API; fundamentally, any rendering
algorithm can be encapsulated behind ANARI; the application developer provides
the aforementioned, structured scene description, periodically updates it as
required, and calls dedicated API functions to render images. These images are
later available in dedicated memory regions. ANARI also hides from the app
developer if rendering happens on the CPU or the GPU. The ANARI implementation
by Intel for example uses OSPRay internally and is thus well-optimized for
CPUs.  The user is free to choose an implementation optimized for their
platform, such as one of the ANARI implementations by NVIDIA. Other ANARI
implementations exist that are more focused on scientific use cases and
research on rendering algorithms themselves~\cite{visionaray}, i.e., ANARI is
also used by 3D rendering researchers who can directly integrate and benchmark
their developments inside vis apps like ParaView that integrate ANARI in their
front-ends.

In this paper we specifically look at challenges that come with data parallel
rendering where the data is (or must be) distributed across a number of nodes
or GPUs. An example of such data, simulated by NASA~\cite{Jones2019marslander}
is shown in \cref{fig:teaser}. Solutions tackling data parallel rendering under
ANARI have recently been proposed by Wald et al.~\cite{dpanari}. This paper
takes a step back and reflects on the problems leading to those solutions and
tries to convey them to a (visualization) layperson audience who are familiar
with concepts from HPC in general, but not with concepts from scientific
visualization and ray tracing in particular. We also demonstrate a case study
visualizing the lander data set on the H100 GPU partition of the RAMSES
supercomputer, which was installed at the University of Cologne in 2024.

\section{Background}
We assume that the reader of this paper is a student or practitioner in HPC but
has little or only passing knowledge in scientific visualization and rendering.
We first describe relevant aspects of the ray tracing rendering algorithm on a
high level. We then discuss what the specific challenges are if the data that
is rendered is not present on a single node or GPU, but is distributed among
several processors.

\subsection{Ray Tracing}
Ray tracing is one of the classic techniques to render 2D images from a 3D
scene description. Independent of the algorithm, doing so requires one to solve
the visibility problem, i.e., identifying those objects that are in front of
all the other objects when viewed from a certain position. The rasterization
algorithm, which is another classic technique, solves this by using a z-buffer
data structure that stores a depth value per 2D image pixel; objects are
projected from 3D to 2D and then rasterized into the z-buffer; depth pixels
associated with objects that are behind other object are discarded. Ray tracing
instead solves the visibility problem by geometrically intersecting objects
with straight lines; these lines can originate at the virtual camera. More rays
are spawned at intersection positions, allowing to compute reflections,
shadows, etc. For a detailed introduction to ray tracing we refer the reader to
classical text books by Marschner and Shirley~\cite{fundamentals} or by
Glassner~\cite{introduction}.

An important aspect of ray tracing is that it is rather intuitive to implement
effects such as reflection, refraction, shadows, etc.; for example, to compute
if a certain spot on a surface is in shadow, one simply casts a ray towards a
light source---if an object intersection was found between the surface spot and
the light source, the surface is in shadow (with respect to that light source),
and if it is not, the surface must be shaded.

Visibility tests involve testing rays against each object in the scene, making
this an inherently inefficient algorithm if not optimized properly. Any
ray tracing library will use some kind of acceleration structure allowing to
cull a majority of objects that are not close to the trajectory of the ray;
different strategies exist, but the most popular and successful one used today
is the bounding volume hierarchy (BVH) (e.g.,~\cite{wald:2007}). The main idea
is to compute a bounding object (usually an axis-aligned box, AABB) around a
set of objects in the scene; by first testing the ray against the bounding
object, which is cheap, we can determine if we possibly hit any of the objects
inside. Only if the bounding object was hit do we need to intersect with
individual objects; these individual objects can, however, themselves be
smaller bounding objects, which gives rise to hierarchical data structures
(usually trees) where the actual geometric objects (surfaces, volume elements,
etc.) are at the leaf nodes. Rays are traversed down those trees until they
find a leaf. Then, usually a handful of surfaces are intersected. If no surface
was hit, traversal continues resursively through the tree.

Implementing efficient BVHs---ones that are fast to traverse, i.e., few
traversal steps down the tree and few surface tests until an intersection was
found, if any---is quite challenging and requires many low-level optimizations.
These optimizations differ significantly between CPU and GPU and are the subject of
state-of-the-art research even today~\cite{vaidynanathan,ylitie}. Libraries
like Embree~\cite{embree} and OptiX~\cite{optix} implement what is the current
state-of-the-art regarding these acceleration structures. GPU architectures
like NVIDIA's RTX even have dedicated chips to accelerate ray-BVH traversal.

\subsection{Data Parallel Ray Tracing}
\begin{figure}[tb]
\begin{center}
  \includegraphics[width=.7\columnwidth]{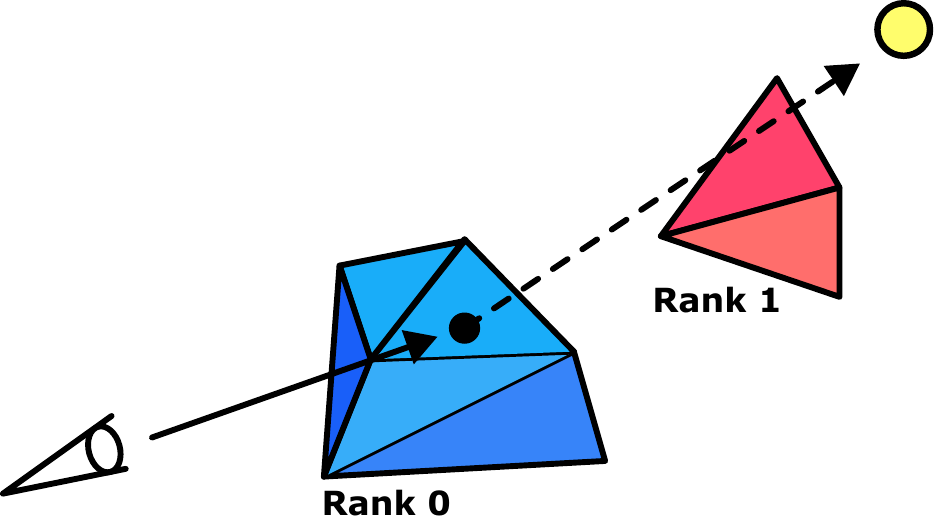}
\end{center}
\vspace{-1em}
\caption{Scenario that presents a challenge for data parallel ray tracing.
  The data (tetraedra, finite elements, etc.) is distributed across multiple
  MPI ranks. Here, a ray was traced from the camera and intersects with the data on
  rank~0. A shadow ray is cast to determine if the intersection point requires
  shading. There is an occluder, but on rank!1, so that the secondary
  ray needs to be sent there first before the operation can complete.
  \label{fig:shadow}
  \vspace{-1em}
}
\end{figure}
Data parallel ray tracing is typically implemented by spatially distributing
the data across ranks or processes. If the data for example consists of finite
elements~\cite{star} these elements need to be distributed so that each rank
gets a share of them. The distribution can be very simple, e.g., so that each
rank gets an equal share. But a common scenario where shading is complex
requires that rays are exchanged by ranks to test for visibility and occlusion.
This is illustrated in \cref{fig:shadow}. Here the data shown in blue is
located on (MPI) rank~0. A ray was traced from the camera and generated a hit
on rank~0. We now need to determine if the hit point is in shadow. The light
source is occluded by the data on rank~1 (red elements); the shadow ray sent to
determine if the hit point is occluded must be traced on that rank. We do not
know in advance where such hits occur and must hence make sure to test the rays
on each rank.

To implement this, rays must be queued and batched up because sending
individual rays would incur prohibitive latency. Wavefront ray tracers will
trace each batch of rays per pixel, then generate a next batch per
``bounce''---i.e., all ranks now process shadows, reflections, etc. A
traditional school of techniques advocated that the number of rays in those
batches is reduced by using culling data structures such as
kd-trees~\cite{zellmann-egpgv-2020}. The downside of this is that kd-trees
require significant pre-processing and the algorithms are relatively
complicated and do not generalize easily. Wald et al.~\cite{rqs} recently
showed that on systems with low-latency interconnects (here demonstrated using
NVIDIA NVLink) such culling is not necessary and it is feasible these days to
just trace and cycle every ray on every process. The advantage of this is that
now the data distribution can be arbitrary and potentially just reuse the
exact distribution to ranks that the simulation code used as well.

It is important to note that data parallel ray tracing does not scale well:
adding more processes will not result in lower execution times; if more
processors are available than can fit the data, it is beneficial to combine
data parallel rendering with data replication using hybrid rendering
techniques~\cite{islands}.

\section{Data Parallel Rendering with ANARI (DP-ANARI)}
As stated above, 3D rendering is a well understood task, and though implemented
in a myriad of ways by different apps, efforts like ANARI~\cite{anari} seek to
streamline and consolidate these tasks under common APIs. We refer
to~\cite{anari} for a detailed description of the Khronos standard ANARI and
its API and only summarize what is important here. ANARI's core data structure
is a render graph: geometry or volumes nodes, materials, light sources,
transformations, and others, form a hierarchy of at most two levels.  A
complete render graph is referred to as a \code{ANARIWorld}. Worlds describe
the virtual scene, and objects in the world (as well as the world itself) are
reference-counted. To render a world, the user creates a \code{ANARIFrame}
object, a \code{ANARICamera}, and a \code{ANARIRenderer}. World, camera, and
renderer become child objects of the frame, which serves as a virtual film abstraction
that also has a memory buffer that can hold the final image pixels.

Notably, ANARI is a \emph{not} a data parallel API; the standard does not
mention anywhere that apps can run in a data distributed way. Wald
et al.~\cite{dpanari} propose a set of conventions for the vis app developer
to follow so that it is compatible with data parallel ANARI implementations.
The paper assumes the use of MPI for this. An important assumption is that the
rendering implementation uses ray queue cycling~\cite{rqs} so the data does not
need to be redistributed but can directly come from the simulation.
Redistribution is not necessary because every ray will eventually be cycled to
every compute node, rendering complicated culling with kd-trees and the like
unnecessary.

The conventions this paradigm (called DP-ANARI by the authors) imposes include
that certain operations are collective similar to how MPI has collective
operations; examples of that are calls to \code{anariRender()} to retire image
pixels into the framebuffer using ray tracing. The framebuffer itself can be
distributed but will eventually be available on the main MPI rank; in contrast,
the scene represented by the \code{ANARIWorld} is assumed to be distributed and
updates by the worker ranks happen locally. The app itself is assumed to
perform MPI synchronization between API calls.

\section{Interactive In-Situ Visualization on RAMSES Using DP-ANARI}
\begin{figure}[tb]
\begin{center}
  \includegraphics[width=.96\columnwidth]{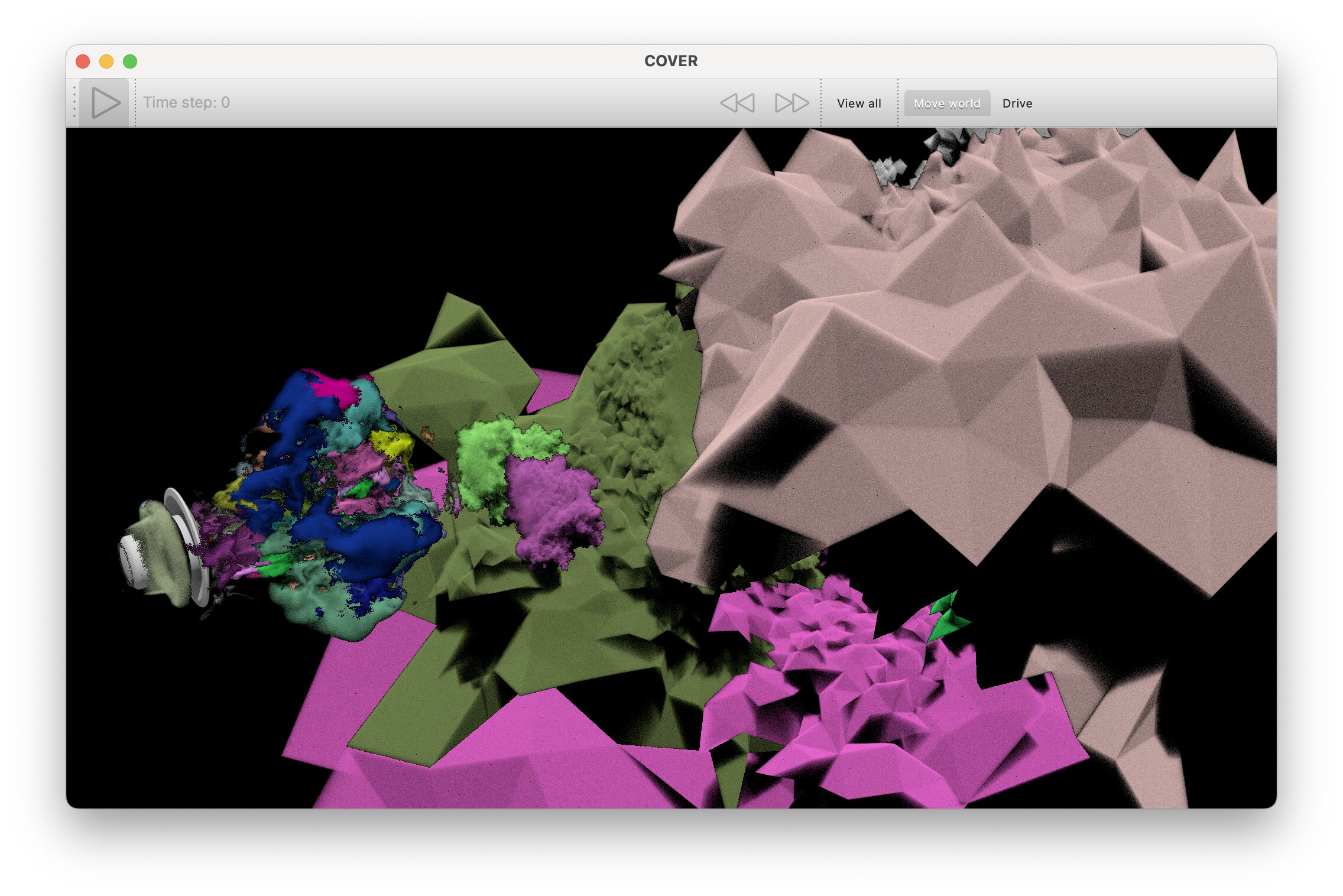}
\end{center}
\vspace{-2em}
\caption{Mars Lander data set with color coding indicating which finite elements
  go on which rank.
  \label{fig:lander-per-rank}
  \vspace{-1em}
}
\end{figure}

DP-ANARI is used for large-scale rendering on the supercomputer RAMSES at the
University of Cologne. One of the visualizations benefiting from this is the
Mars Lander seen in \cref{fig:teaser} that is freely available as simulation
data from NASA~\cite{Jones2019marslander}. A speciality of that data set is
that it is available in the exact way the data was distributed to MPI ranks
during the simulation. We refer the reader to the paper by Sahistan et
al.~\cite{sahistan:2024} for a detailed overview of this mixed finite element
data set. The challenges with this data set are its size, and the fact that the
data is distributed unevenly in space (see \cref{fig:lander-per-rank}). It is
hence well-suited to DP-ANARI and ray queue cycling with its low
sensitivity to spatially unbalanced object distribution. The fact that the data
is distributed in the exact way it was simulated on Summit allows us to
replay the simulation and perform the visualization as if it was running
in-situ.

\begin{figure}[tb]
\begin{center}
  \includegraphics[width=.96\columnwidth]{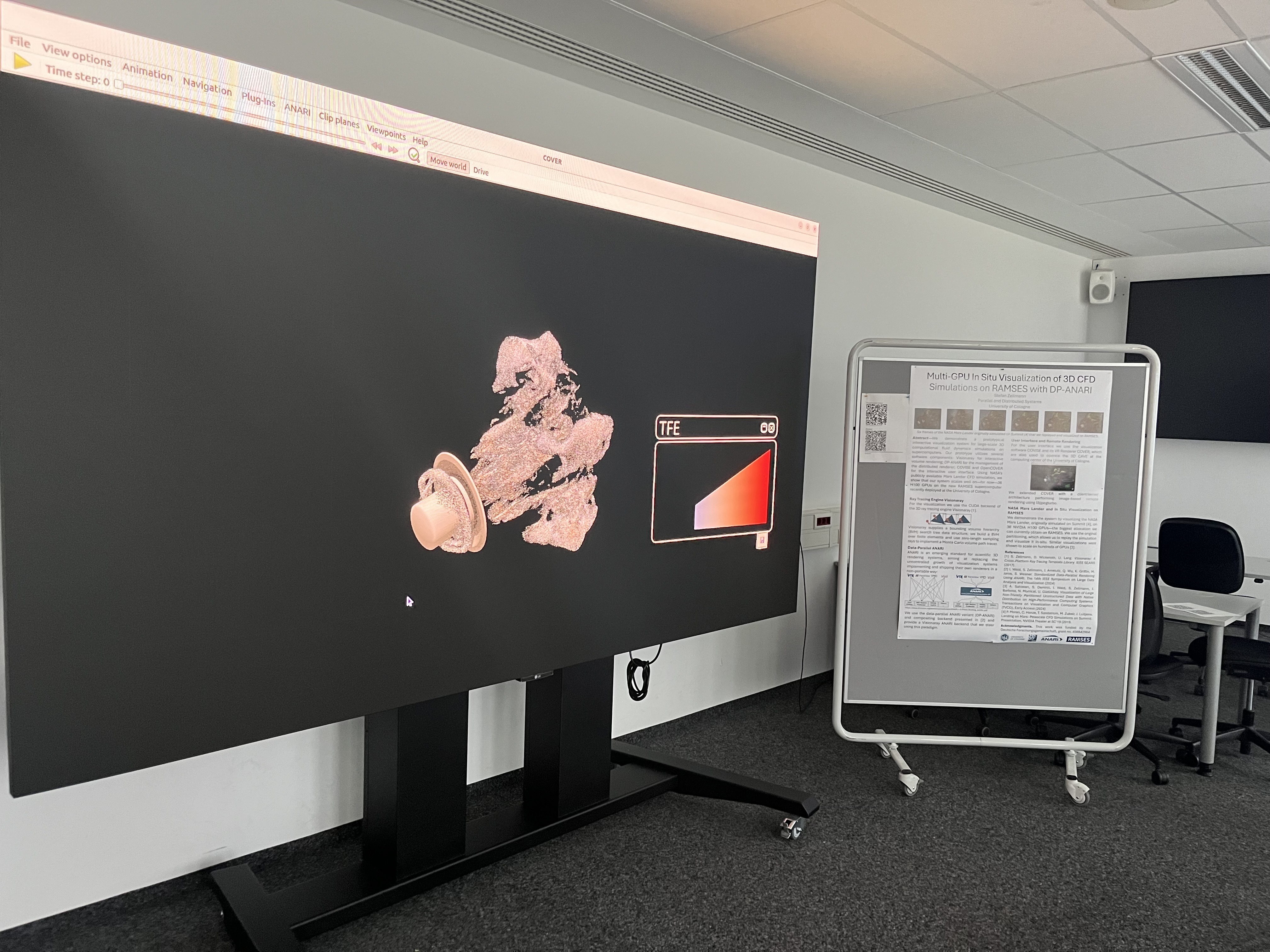}
\end{center}
\vspace{-1em}
\caption{Thin display client connecting to DP-ANARI distributed
  MPI renderer on Cologne's super computer RAMSES, here operating a display
  wall while RAMSES renders the NASA Mars Lander on 36 NVIDIA~H100 GPUs.
  \label{fig:display}
  \vspace{-1em}
}
\end{figure}
One of the challenges on modern HPC systems is that OpenGL is often not
available on compute nodes; on RAMSES, e.g., the users are not allowed to
start their own \code{Xorg} instances, prohibiting the use of ParaView and
\code{pvserver}, ParaView's MPI server instance; though DP-ANARI does not need
OpenGL, \code{pvserver} does so (although we bypass its internal compositing).
We instead opted to integrate DP-ANARI into the virtual reality (VR) renderer
OpenCOVER developed at HLRS~\cite{covise}. OpenCOVER can also be run in MPI
cluster mode but allows the user to configure that the processes run headless
and do not require a local OpenGL context to render and composite images. We
added ANARI with the DP-ANARI extensions as an OpenCOVER plugin; we also added
a remote rendering server operated by the plugin that allows one to connect to
the OpenCOVER instance running on RAMSES with another OpenCOVER instance acting
as a thin display server. An example of this thin client running the Mars
Lander simulation can be seen in \cref{fig:display} where it is used to operate
an LED display.

The bigger challenges with this data set are its size---as we have only 36~GPUs
available on RAMSES (yet with much more memory than the GPUs on Summit had) we
cannot directly emulate a simulation rerun, but must use multiple instances per
GPU. The pragmatic way of doing so was to just trivially recombine the data so
we do not have to run 500 MPI processes on 36~GPUs. Another challenge is the
one mentioned above with occluders potentially ending up on different nodes
than the finite elements that were initially hit, as illustrated in
\cref{fig:shadow}. This can also be seen in \cref{fig:lander-per-rank}, which
shows the lander data set color coded by rank assignment.

For rendering we can currently use the two implementations proposed by Wald et
al.~\cite{dpanari}, which are available as open source projects.
\emph{Barney}~\cite{barney} is a full-fledge MPI data parallel renderer with a
low level interface and ANARI front-end implementing the ray queue cycling
paradigm using NVIDIA~OptiX.

We were able to fit a significant portion of the Lander data set onto the 36
H100 GPUs. As the data is time-varying we follow a strategy to selectively load
and later evict time steps so we do not initially hold the whole data set in
GPU memory, but only a subset of the time steps. We achieve interactive frame
rates (usually 10-15 frames per second, sometimes lower, e.g., when moving the
virtual camera into the data set). We note that OpenCOVER is also used to
operate the CAVE virtual reality environments at both HLRS in Stuttgart and the
IT-center (ITCC) at the University of Cologne, so that in future work we plan
to use the newly developed ANARI plugin for in-situ visualization in virtual
reality.

\section{Conclusion}
We presented data parallel rendering on the new RAMSES supercomputer at the
University of Cologne, using a case study visualizing the NASA Mars Lander data
set in-situ. The focus of the paper was also to convey some of the challenges
with data parallel rendering to a layperson audience with passing knowledge of
scientific visualization and rendering. Data parallel renderers use ray tracing
these days, which makes computing secondary effects like shadows or reflections
very intuitive. This, however, requires more complex communication patterns
such as wavefronts and ray queue cycling than the patterns used when not
simulating these effects. We demonstrated an efficient implementation of this
DP-ANARI paradigm that we integrated into the open source renderer OpenCOVER,
which is jointly developed and used by HLRS and the University of Cologne.

\section*{Acknowledgments}
This work was supported by the Deutsche Forschungsgemeinschaft (DFG, German
Research Foundation) under grant no.~456842964. We express our gratitude to the
IT Center of the University of Cologne (ITCC) who supported this work by
graciously provided us with computing resources on the supercomputer RAMSES. We
also thank the HPC team of the University of Cologne for their support in
running our code on RAMSES.

%-------------------------------------------------------------------------
% bibtex
\bibliographystyle{eg-alpha-doi}  
\bibliography{main}

\end{document}